\documentclass[twocolumn,showpacs,prl,amsfonts,amsmath,amssymb,floatfix]{revtex4} 
\usepackage{color}
\usepackage{hhline}	
\usepackage{mathrsfs}
\usepackage{graphicx}
\usepackage{dcolumn}
\usepackage{bm}
\usepackage{multirow}
\usepackage{booktabs}
\usepackage{afterpage}
\usepackage{amsmath}

\begin{document}

\title{Correlated Band Structure of a Transition Metal Oxide ZnO Obtained from a Many-Body Wave Function Theory}

\author{Masayuki Ochi$^{1,2}$}
\author{Ryotaro Arita$^{2}$}
\author{Shinji Tsuneyuki$^{3,4}$}
\affiliation{$^1$Department of Physics, Osaka University, Machikaneyama-cho, Toyonaka, Osaka 560-0043, Japan}
\affiliation{$^2$RIKEN Center for Emergent Matter Science (CEMS), Wako, Saitama 351-0198, Japan}
\affiliation{$^3$Department of Physics, The University of Tokyo, Hongo, Bunkyo-ku, Tokyo 113-0033, Japan}
\affiliation{$^4$Institute for Solid State Physics, The University of Tokyo, Kashiwa, Chiba 277-8581, Japan}

\date{\today}
\begin{abstract}
Obtaining accurate band structures of correlated solids has been one of the most important and challenging problems in first-principles electronic structure calculation. There have been promising recent active developments of wave function theory for condensed matter, but its application to band-structure calculation remains computationally expensive. In this Letter, we report the first application of the bi-orthogonal transcorrelated (BiTC) method: self-consistent, free from adjustable parameters, and systematically improvable many-body wave function theory, to solid-state calculations with $d$ electrons: wurtzite ZnO.
We find that the BiTC band structure better reproduces the experimental values of the gaps between the bands with different characters than several other conventional methods.
This study paves the way for reliable first-principles calculations of the properties of strongly correlated materials.
\end{abstract}

\maketitle

To reveal fertile and nontrivial physics in condensed matter, first-principles electronic-structure calculation has established itself as an indispensable tool in recent studies.
For this purpose, density functional theory (DFT)~\cite{HK, KS} has played a leading role and has been applied to various materials; however, the limitations of this theoretical framework have come to light.
One of the major problems is an inaccurate description of strong electron correlations, e.g., in transition metal oxides.
The $GW$ method~\cite{GW1,GW2,GW3} is a promising way to ameliorate the inaccuracy of the band structures and has been applied to several solids, including $d$-electron systems.
However, because the $GW$ method is often applied without satisfying self-consistency, a nontrivial dependence on the initial DFT calculations is introduced.
It has also been reported that the $GW$ method sometimes exhibits severe difficulty in obtaining converged results~\cite{GWconvergence}, owing to its perturbative nature.
Another possible choice is to construct effective models from DFT that include correlation terms such as Hubbard $U$ and to solve the models using elaborated methodologies~\cite{Models} such as dynamical mean-field theory~\cite{DMFTreview,DMFTreview2,DMFT2}. However, the exact correspondence between the effective models and the first-principles Hamiltonian is a nontrivial problem.

Recently, wave function theory (WFT), which had been mainly applied to molecular systems and established itself as the gold standard in theoretical chemistry~\cite{Szabo}, has become a promising alternative to DFT for accurate descriptions of electron correlation in solids~\cite{FCIQMC}.
Among the most powerful frameworks in WFT are first-principles quantum Monte Carlo (QMC) methods~\cite{QMCreview,QMCreview2}, such as the variational Monte Carlo (VMC), diffusion Monte Carlo (DMC), auxiliary-field quantum Monte Carlo (AFQMC), and full-configuration-interaction (FCI) QMC methods.
Other kinds of WFT, called post-Hartree-Fock (post-HF) methods, have also been applied to condensed matter in recent years~\cite{MP2_Kresse_band,MP2_Kresse_totE,MP2_Shepherd,CCD_Shepherd,CCD_Shepherd2}.
However, their targets are in most cases limited to solids with small unit cells, owing to their expensive computational cost.
In addition, the correlated band structure, which is quite useful in various kinds of theoretical analyses, is not easily obtained in many WFTs.
For example, calculation of the band structure in the framework of VMC or DMC requires a large number of single-point calculations of the excited states, which is a clear difference from a mean-field-like approach such as DFT, whereby the whole band structure is obtained at once.

From this viewpoint, the transcorrelated (TC) method~\cite{BoysHandy, Handy, Ten-no, Umezawa} is a fascinating WFT that can be applied to solids with reliable accuracy and moderate computational cost~\cite{Sakuma, TCaccel, TCjfo, TCCIS, TCMP2, TCPW}.
The TC method adopts the so-called Jastrow ansatz, which is based on a promising strategy often adopted in several WFTs such as QMC methods to describe strong electron correlations; i.e., the electron-electron distance is included into many-body wave functions. Explicitly correlated electronic-structure theory~\cite{R12} in quantum chemistry also adopts this strategy.
In fact, the Gutzwiller- and Jastrow-correlation factors have often been used to describe strong electron correlation, including the Mott physics in systems such as the Hubbard model~\cite{VMC,MottVMC1,MottVMC2,MottVMC3}.
It is also important to note that, unlike several WFTs, the whole band structure is obtained at once by solving a one-body self-consistent-field (SCF) equation in the TC method, as described later in this Letter. Moreover, the TC method is deterministic, i.e., free from the statistical error, unlike the QMC methods.
Accurate calculations for the Hubbard model~\cite{TCHubbard,LieAlgebra} and molecular systems~\cite{LuoTC,LuoVTC,CanonicalTC} were also reported using the TC method or other theories that have a close relationship with the TC method.
However, insofar as solid-state calculations are concerned, the TC method has so far been applied only to weakly correlated systems.

In this Letter, we present the first application of the TC method to the band-structure calculation of a $d$-electron system: wurtzite ZnO. 
$3d$ transition metal oxides have posed theoretical challenges for first-principles band-structure calculations,
as it is well known that popular approximations such as the local-density approximation (LDA) fail to provide their band structures accurately, as we shall see later.
We find that the TC method with the bi-orthogonal formulation (the BiTC method)~\cite{BiTC,TCMP2} successfully reproduces the experimental band structure of ZnO. 
We also clarify how the Jastrow factor improves the first-principles description of the correlated electronic states through comparison of the band structures and electron densities among the BiTC and other methods.

The central concept of the TC method is to make use of the similarity transformation of the many-body Hamiltonian with the Jastrow factor $F=\mathrm{exp}(-\sum_{i,j} u(x_i,x_j))$,
\begin{align}
\mathcal{H}\Psi = E\Psi \Leftrightarrow \mathcal{H}_{\mathrm{TC}}\Phi = E \Phi\ (\mathcal{H}_{\mathrm{TC}} =F^{-1}\mathcal{H}F), \label{eq1}
\end{align}
where the correlated wave function is represented as $\Psi=F\Phi$ and $\mathcal{H}_{\mathrm{TC}}$ is called the TC Hamiltonian.
Here, $x$ denotes a pair of space and spin coordinates: $x=(\mathbf{r}, \sigma)$.
By adopting the so-called Slater-Jastrow ansatz, $\Phi$ becomes a Slater determinant consisting of one-electron orbitals, $\phi(x)$: $\Phi=\mathrm{det}[\phi_i(x_j)]$, and Eq.~(\ref{eq1}) yields an SCF equation for one-electron orbitals that experience the effective interaction described with the TC Hamiltonian:
\begin{align}
&\left( -\frac{1}{2}\nabla_1^2 +v_{\mathrm{ext}}(x_1) \right) \phi_i (x_1)\notag \\
&+ \sum_{j=1}^N
\int \mathrm{d}x_2\  \phi_j^*(x_2) v_{\mathrm{2body}}(x_1,x_2)
\mathrm{det} \left[
\begin{array}{rrr}
\phi_i(x_1) & \phi_i(x_2) \\
\phi_j(x_1) & \phi_j(x_2) \\
\end{array} \right] \notag \\
&- \frac{1}{2}\sum_{j=1}^N \sum_{k=1}^N
\int \mathrm{d}x_2 \mathrm{d}x_3\  \phi_j^*(x_2)\phi_k^*(x_3)v_{\mathrm{3body}}(x_1,x_2,x_3)  \notag \\
&\times 
\mathrm{det} \left[
\begin{array}{rrr}
\phi_i(x_1) & \phi_i(x_2) &  \phi_i(x_3) \\
\phi_j(x_1) & \phi_j(x_2) & \phi_j(x_3) \\
\phi_k(x_1) & \phi_k(x_2) & \phi_k(x_3)
\end{array} \right]
= \sum_{j=1}^N \epsilon_{ij} \phi_j(x_1), \label{eq:SCF}
\end{align}
where $v_{\mathrm{ext}}(x_1)$, $v_{\mathrm{2body}}(x_1,x_2)$, and $v_{\mathrm{3body}}(x_1,x_2,x_3)$ are the external potential including the nucleus-electron interaction~\cite{note_Vext} and the two- and three-body effective interactions in the TC Hamiltonian, defined as
\begin{align}
&v_{\mathrm{2body}}(x_1,x_2)\equiv \frac{1}{|\mathbf{r}_1-\mathbf{r}_2|}+\frac{1}{2}\sum_{i=1}^2 [ \nabla_i^2 u(x_1,x_2)\notag\\
&-(\nabla_i u(x_1,x_2))^2+ 2\nabla_i u(x_1,x_2)\cdot \nabla_i],\label{eq:2body}
\end{align}
and
\begin{align}
&v_{\mathrm{3body}}(x_1,x_2,x_3)
\equiv\nabla_1 u(x_1,x_2)\cdot \nabla_1 u(x_1,x_3) \notag\\
&+ \nabla_2 u(x_2,x_1) \cdot \nabla_2 u(x_2,x_3)
+ \nabla_3 u(x_3,x_1) \cdot \nabla_3 u(x_3,x_2),
\end{align}
respectively. As is evident, the HF method can be regarded as the TC method with $u=0$. 
Owing to this effective one-body picture, it is possible to treat the many-body correlation with moderate computational cost.
In addition, one can obtain the band structure of the quasiparticles by using the real part of the eigenvalues of the $\epsilon$ matrix on the right-hand side of Eq.~(\ref{eq:SCF}).
One of the authors proved in prior work~\cite{Umezawa} that such a use of the $\epsilon$ matrix as quasiparticle energies is consistent with Koopmans' theorem.
We note that one can systematically improve the accuracy of the TC method by utilizing quantum chemical methodologies such as the coupled-cluster and configuration interaction methods to go beyond a single Slater determinant (e.g. Refs.~\cite{Ten-no, BiTC, TCMP2}).

Here, the Jastrow function, $u(x,x')$, is set to the following simple form without adjustable parameters:~\cite{Sakuma,QMCreview,Ceperley,CeperleyAlder}
\begin{equation}
u(x,x')=\frac{A}{|\mathbf{r}-\mathbf{r'}|}
\left( 1-\mathrm{exp}\left( -|\mathbf{r}-\mathbf{r'}|/C_{\sigma,\sigma'} \right) \right) ,
\label{eq:Jastrow}
\end{equation}
where $A=\sqrt{V/(4\pi N)}$ ($N$ is the number of valence electrons in the simulation cell, $V$ is the volume of the simulation cell) and $C_{\sigma, \sigma'} = \sqrt{2A}$ (spin parallel:$\ \sigma=\sigma'$), $\sqrt{A}$ (spin antiparallel:$\ \sigma\neq\sigma'$).
The long-range asymptotic form of this function describes the screening effect of the electron-electron Coulomb interaction~\cite{BohmPines}.
The short-range behavior of the exact Jastrow function should obey the cusp condition~\cite{cusp,cusp2,cusp_note}.
The Jastrow ansatz adopted here works well for state-of-the-art QMC methods~\cite{QMCreview,QMCreview2}.
Although our choice of the Jastrow function is rather simple,
we shall see that, nevertheless, it works well not only for weakly correlated systems~\cite{Sakuma,TCaccel} but also for the 3$d$-electron system.
Of course, it is possible to improve a quality of the many-body wave function by using a complicated Jastrow factor, but we adopted this simple trial wave function to realize moderate computational cost.

In this study~\cite{calc_note}, we adopted the BiTC method~\cite{BiTC,TCMP2}, in which the left one-electron orbitals, $\chi(x)$ in the left Slater determinant X $=\mathrm{det}[\chi_i(x_j)]$, replace the bra orbitals in the SCF equation~(\ref{eq:SCF}), while the ket orbitals remain $\phi(x)$.
Because the TC Hamiltonian is non-Hermitian, $\phi(x)$ and $\chi(x)$ become different.
We do not show the band structure calculated using the TC method without the bi-orthogonal extension here, because of the large imaginary part of the eigenvalues~\cite{Sppl}.

\begin{figure}
 \begin{center}
  \includegraphics[width=8.6cm]{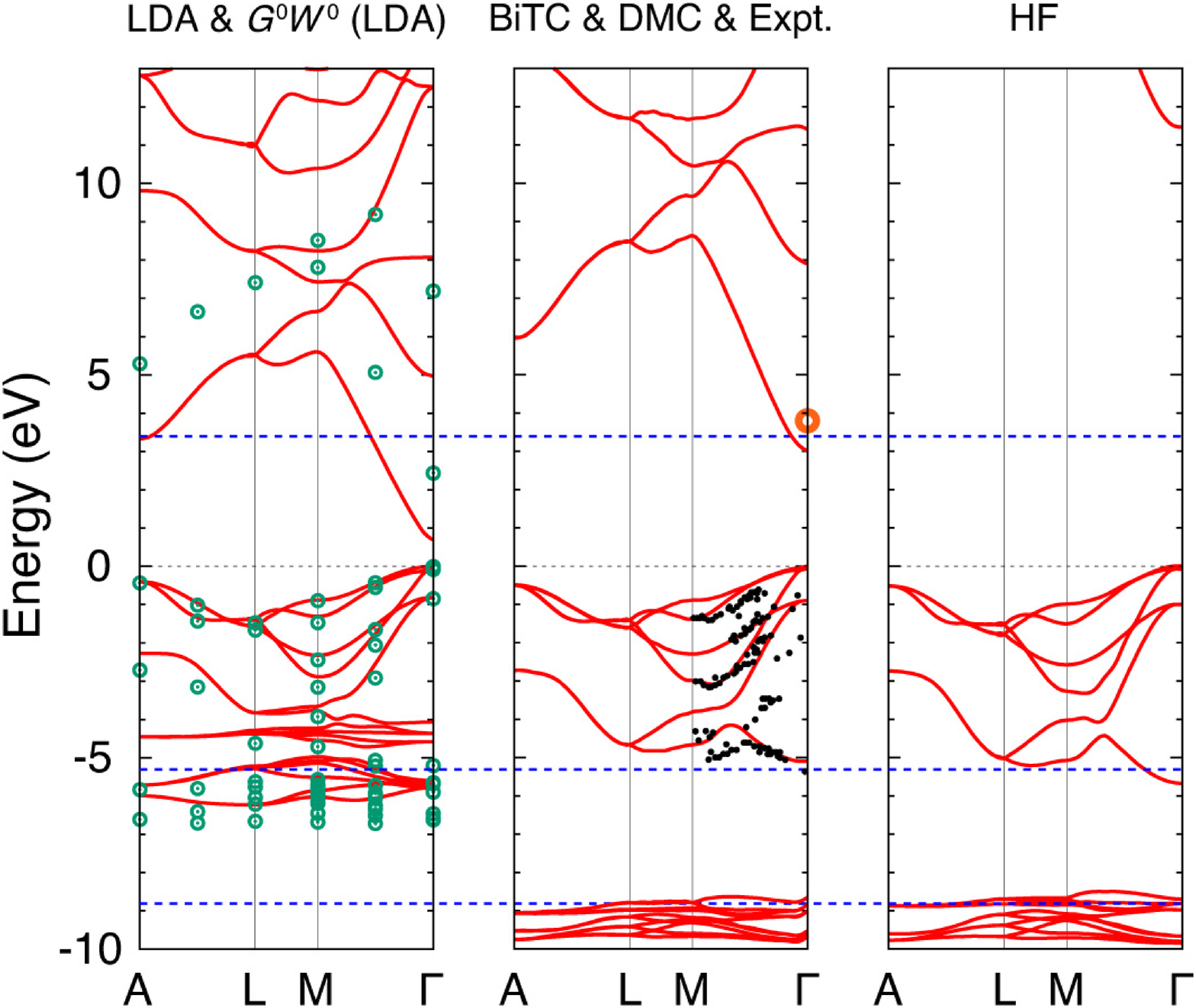}
  \caption{Calculated band structures using the LDA (solid lines), all-electron $G^0W^0$~\cite{GWZnO} (green circles), BiTC (solid line), DMC~\cite{ZnO_DMC} (orange circles), and HF (solid line) methods. Experimental data taken from Ref.~\cite{ZnOexpt_dots} are shown with black dots in the middle figure. Blue broken lines show the other experimental data for the positions of the conduction-band minimum, the O-$2p$ bottom~\cite{ZnOexpt_old}, and the Zn-$3d$ peak position~\cite{ZnO3d}, which might correspond to the averaged position of the Zn-$3d$ bands. The valence-band maximum energy is set to zero.}
  \label{fig:1}
 \end{center}
\end{figure}

\begin{table*}
\begin{center}
\footnotesize{\tabcolsep = 0.6mm
\begin{tabular}{c c c c c c c c}
\hline \hline
 & & \multicolumn{2}{c}{Band gap [Error]} &  \multicolumn{2}{c}{O-$2p$ bottom [Error]} & Zn-$3d$ averaged & Zn-$3d$ bottom \\
 \hline
DFT & LDA & 0.7 & [$-$2.7] &  $-$ & $-$ & $-$ & $-$5.8  \\
 & HSE03$^a$& 2.1 & [$-$1.3] & $-$4.9 & [$+$0.4] & $-$ & $-$6.5 \\ 
$GW$ & $G^0W^0$(LDA)$^b$ & 2.4 & [$-$1.0] &  $-$5.2 & [$+$0.1] & $-$  & $-$6.5 \\ 
& $G^0W^0$(HSE03) & 3.2$^a$, 3.46$^c$ & [$-$0.2, $+$0.06] & $-$ & $-$ &$-$6.21$^c$ & $-$7.2$^a$  \\ 
& $G^0W^0$ (GGA+$U$)$^c$ & 2.94 & [$-$0.46] & $-$5.6 & [$-$0.3] & $-$6.33 & $-$7.1  \\ 
 & ($U-J=$ 6 eV) & & & & & & \\
& $G^0W^0$+$V_d$ (GGA+$U$)$^c$ & 3.30 & [$-$0.1] & $-$5.5 & [$-$0.2] & $-$7.45 & $-$8.0  \\ 
 & ($U-J=$ 6 eV, $V_d=$ 1.5 eV) & & & & & & \\
& QS$GW$$^d$ & 3.87 & [$+$0.47] &$-$5.3 & [$\pm$ 0] & $-$  & $-$7.2 \\ 
& sc$GW$ (RPA) $^e$ & 3.8 & [$+$0.4] & $-$ & $-$ & $-$6.4 & $-$  \\ 
& sc$GW$ ($e$-$h$)$^e$ & 3.2 & [$-$0.2] & $-$ & $-$ & $-$6.7 & $-$  \\ 
WFT & AFQMC$^f$ &  3.26(16) & [$-$0.14] &$-$ & $-$ &$-$&$-$ \\ 
& DMC$^g$ & 3.8(2) & [$+$0.4] & $-$ & $-$ &$-$&$-$ \\ 
& BiTC & 3.1 & [$-$0.3] & $-$5.1 & [$+$0.2] & $-$9.1 & $-$9.7  \\
& HF& 11.4 & [$+$8.0] & $-$5.7 & [$-$0.4] & $-$9.3 & $-$9.9  \\
Expt.&  & \multicolumn{2}{c}{3.4$^h$} &\multicolumn{2}{c}{$-$5.3$^h$, $-$5.2(3)$^i$} &$-$7.5$^c$, $-$7.5(2)$^j$, $-$8.5(4)$^k$,  & $-$\\
 & & & & & &  $-$8.6(2)$^k$, $-$8.81(15)$^i$ & \\
\hline \hline
\end{tabular}
}
\end{center}
\caption{\label{table:band} Some characteristic values in the band structures of ZnO, as calculated by several methods~\cite{Tab1_note}. The bottoms of the Zn-$3d$ and O-$2p$ bands are evaluated at the $\Gamma$ point where the valence-band top is set to zero for each method. The errors between the calculated and experimental values shown in Ref.~\cite{ZnOexpt_old} are presented in parentheses.
For LDA, the O-$2p$ bottom and Zn-$3d$-averaged levels are not presented here because of the overlap among the O-$2p$ and Zn-$3d$ bands.
For $G^0W^0$ (HSE03), the Zn-$3d$ bottom position was read from the density of states presented in Ref.~\cite{ZnOHSEGW}.
All values are in eV.
$^a$ Ref.~\cite{ZnOHSEGW}. $^b$ Ref.~\cite{GWZnO}. $^c$ Ref.~\cite{ZnOARPES}. $^d$ Ref.~\cite{ZnOQSGW}. $^e$ Ref.~\cite{ZnOscGW}. $^f$ Ref.~\cite{ZnOAFQMC}. $^g$ Ref.~\cite{ZnO_DMC}. $^h$ Ref.~\cite{ZnOexpt_old}. $^i$ Ref.~\cite{ZnO3d}. $^j$ Ref.~\cite{ZnO3d_3}. $^k$ Ref.~\cite{ZnO3d_2}.  }
\end{table*}

Figure~\ref{fig:1} presents the band structures of ZnO calculated with the LDA, all-electron $G^0W^0$ starting from LDA using the LAPW~\cite{LAPW} method, BiTC, and HF methods.
The characteristic energy values in these band structures, as evaluated at the $\Gamma$ point, are listed in Table~\ref{table:band}.
Table~\ref{table:band} also lists calculated values with various other methods.
By focusing on the gaps between the bands with different characters as listed in Table~\ref{table:band}, the BiTC band structure exhibits better accuracy than many other conventional methods, including the $G^0W^0$ method starting from LDA, and accuracy comparable to that for the most accurate varieties of the $GW$ scheme, i.e., the $G^0W^0$ method starting from HSE03 and the self-consistent $GW$ methods such as the QS$GW$ method~\cite{QSGW}.
It is noteworthy that both the $GW$ and BiTC methods yield such successful results despite being based on conceptually different formulations.
It is important that the BiTC method is based on the self-consistent formulation and thus is independent of DFT calculations, whereas the $G^0W^0$ method strongly depends upon the unperturbed DFT calculations, as seen in Table~\ref{table:band}.
We should also note that, while some methods employ parameters $U$ and $V_d$, which are difficult to determine in the {\it ab initio} way, the BiTC method does not use such parameters.

Consistency of the calculated band gaps among WFTs that use similar trial wave functions (AFQMC, DMC, and BiTC) is also remarkable.
We again stress that the whole band structure is not easily obtained in QMC simulations and requires many single-point excited-state calculations.
We can see that the band structures calculated with the BiTC and HF methods are very similar, except for the band gap.
Therefore, the main role of the Jastrow factor used in this study on the band structure seems to be improvement of the size of the band gap through the screening effect of the electron--electron interaction, which is described with the long-ranged asymptotic behavior of the Jastrow factor.
To obtain a more accurate band structure, e.g., with respect to the depth of the Zn-3$d$ bands, more elaborated Jastrow factors, as used in QMC studies~\cite{QMCreview,QMCreview2}, will be necessary.
Because the Zn-$3d$ bands are almost flat, a key point might be accurate description of the atomic states, which is an important issue for future investigation.

For the $GW$ method, it was pointed out that the shallow Zn-$d$ bands can strengthen the $p$-$d$ hybridization, and thus can result in underestimation of the band gap~\cite{ZnOARPES,ZnOscGW}. A similar situation might also be realized in the BiTC band structure, whereas the Zn-$d$ bands are rather deep~\cite{note_GW} and so the band gap can be overestimated. However, the BiTC band gap is also affected by the $A$ parameter in the Jastrow factor, as mentioned in the above comparison between the BiTC and HF band structures. Because the $A$ parameter used in this study was determined by RPA analysis of the uniform electron gas, it can cause overscreening in the insulator~\cite{TCjfo}, thereby decreasing the band gap. One possibility is that these two factors are canceled here, but more detailed investigation on other materials is also an important future issue~\cite{note_GW2}.

\begin{figure}
 \begin{center}
  \includegraphics[width=8.2cm]{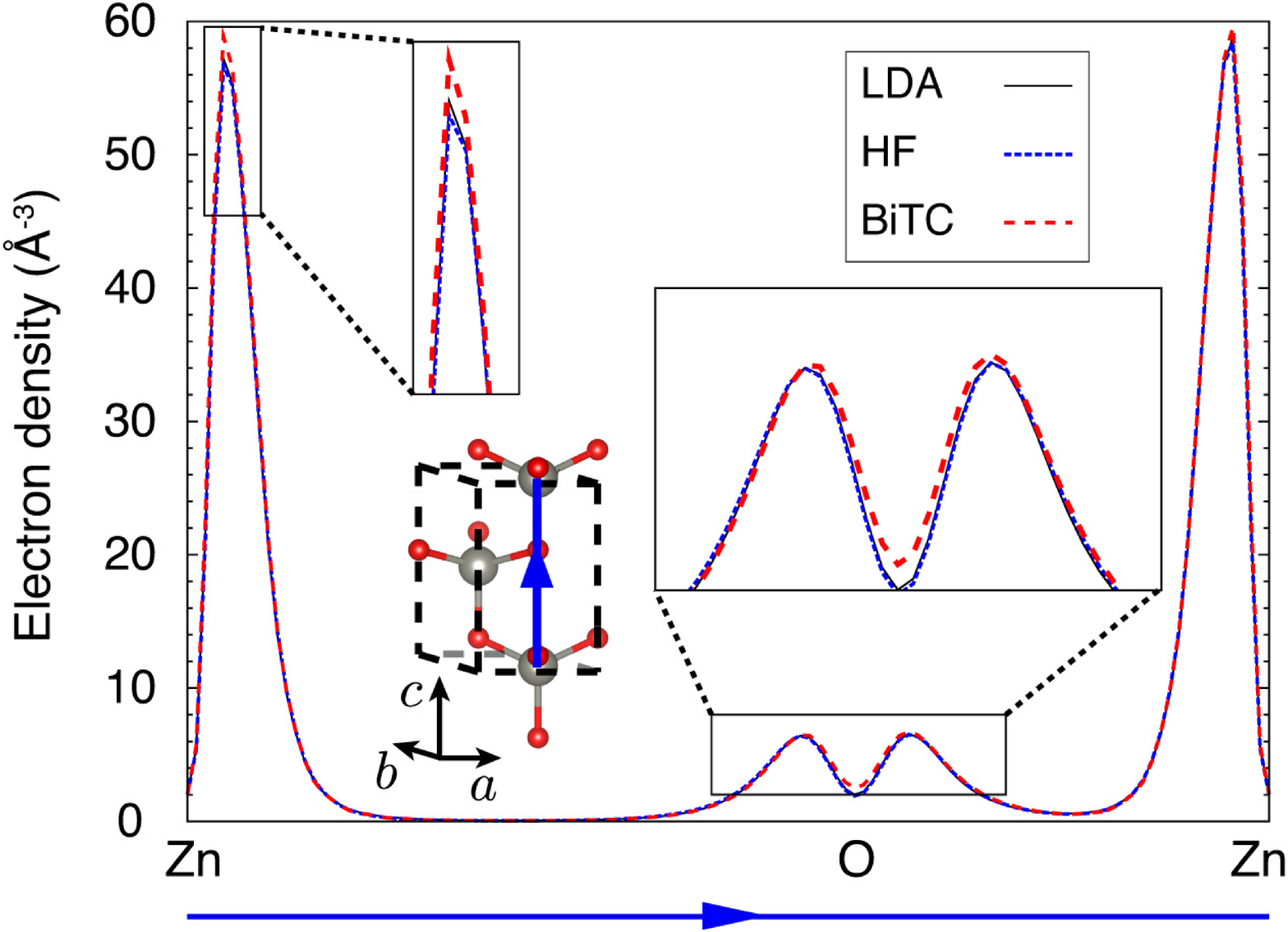}
  \caption{Calculated electron densities on the blue line shown in the crystal structure are presented for the LDA, BiTC, and HF methods with solid black, broken blue, and broken red lines, respectively. The crystal structure was depicted using the \textsc{VESTA} software~\cite{VESTA}.}
   \label{fig:2}
 \end{center}
\end{figure}

Figure~\ref{fig:2} presents the electron densities calculated with the LDA, HF, and BiTC methods on the line shown in the crystal structure.
The electron density obtained by the BiTC method is defined as $n(\mathbf{r})=$ Re$[\sum_{i=1}^N \chi_i^*(\mathbf{r}) \phi_i(\mathbf{r})]$, where the condition 
$\int \mathrm{d}\mathbf{r}\ n(\mathbf{r}) = N$ is satisfied due to the bi-orthonormalization condition $\langle \chi_i | \phi_j \rangle = \delta_{ij}$.
We can see that the electron densities of these methods are almost the same. However, a slight increase of the electron density at the atomic sites is observed for the BiTC method compared with the others.
Such a tendency is consistent with the fact that the strong divergence of the electron-electron Coulomb repulsion is alleviated for the effective two-body interaction in the similarity-transformed TC Hamiltonian, because the Jastrow factor satisfies the cusp condition. 
More concretely, the $\nabla_1^2 u(x_1,x_2)$ and $\nabla_2^2 u(x_1,x_2)$ terms in Eq.~(\ref{eq:2body}) yield $1/|\mathbf{r}_1-\mathbf{r}_2|$ divergence with a different sign than the electron-electron Coulomb repulsion.
As can be seen in the proof of the cusp condition~\cite{cusp,cusp2}, the true many-body wave function should exhibit deformation described with the two-body degrees of freedom near the electron-electron coalescence point, which cannot be represented solely with one-body degrees of freedom.
This is a characteristic advantage of the Slater-Jastrow-type wave function for the description of localized electronic states, such as in strongly correlated systems.
It is noteworthy that the atomic calculations of the TC method also exhibit a similar tendency for localization~\cite{Umezawa}.

Finally, we mention the computational effort required for the BiTC calculation.
Computation takes place on time scales given by $\mathcal{O}(N_k^2 N_b^2 N_{pw} \log N_{pw})$, where $N_k$, $N_b$, and $N_{pw}$ are the numbers of $\mathbf{k}$-points, occupied bands, and plane waves, respectively~\cite{note_computation}. This is the same order as that for the HF or hybrid DFT calculations with a prefactor about 20 to 40~\cite{note_cmptime}.
The BiTC calculation involves neither the frequency index nor the convergence with respect to the number of conduction bands, unlike some perturbative methodologies such as the $GW$ method.
As can be seen from the fact that the hybrid DFT calculations have now been applied to various periodic systems, the computational cost of the BiTC method is reasonable for solid-state calculations.
One remaining obstacle for wide application of the BiTC method is that we use the norm-conserving pseudopotential with a very high cutoff energy to handle the semicore states at present.
However, this problem is not inherent to the BiTC method and can be overcome in principle by the development of a pseudopotential formalism such as the PAW method~\cite{PAW1,PAW2} adapted to the TC method, which is an important future issue~\cite{note_PAW}.
We also note that the deformation of the electron density near atomic sites shown in Fig.~\ref{fig:2} also implies the importance of careful treatment for the core states. This can also be a common problem for QMC calculations using the Jastrow correlation factor. 

To conclude, we apply the bi-orthogonal version of the TC method to wurtzite ZnO and find that it well reproduces the experimental band structure.
Our study encourages further investigation of other strongly correlated materials using the BiTC method.

\begin{acknowledgments}
A part of the calculations was performed using supercomputers at the Supercomputer Center, Institute of Solid State Physics, The University of Tokyo.
This study was supported by Grant-in-Aid for young scientists (B) (No. 15K17724) from the Japan Society for the Promotion of Science, MEXT Element Strategy Initiative to Form Core Research Center, CREST, JST, and Computational Materials Science Initiative, Japan.
\end{acknowledgments}

\clearpage
\onecolumngrid

\section{Supplemental Material}

\setcounter{figure}{0}

\section{Convergence with respect to the number of $\mathbf{k}$-points}

We have performed band-structure calculations using a $2\times 2\times 2$ $\mathbf{k}$-mesh (32 atoms in the unit cell) for the HF and BiTC methods to confirm the convergence with respect to the number of $\mathbf{k}$-points. For the BiTC calculations, the band gap and the relative energy of the Zn-$d$-band bottom to the valence-band top are 2.92 and $-9.73$ eV, respectively, for a $2\times 2\times 2$ $\mathbf{k}$-mesh, whereas they are 3.09 and $-9.68$ eV, respectively, for a $3\times 3\times 3$ $\mathbf{k}$-mesh (108 atoms in the unit cell).
For the HF calculations, the corresponding values are 11.53 and 10.07 eV, respectively, for a $2\times 2\times 2$ $\mathbf{k}$-mesh, whereas they are 11.42 and $-9.90$ eV, respectively, for a $3\times 3\times 3$ $\mathbf{k}$-mesh. The difference between the values calculated using the two $\mathbf{k}$-meshes is less than 0.2 eV for all of these values.

In our previous study on the TC method~\cite{TCaccel}, we saw that the calculated band gap $E_{gap} (N_k)$ with $N_k$ $\mathbf{k}$-points obeys an approximate relation:
\begin{equation}
E_{gap} (N_k) \simeq E_{gap} (N_k=\infty) + C / N_k,
\end{equation}
where $C$ is a constant. Using this approximate relation for the band energies, the residual finite-size error in our calculations using a $3\times 3\times 3$ $\mathbf{k}$-mesh is estimated to be less than 0.1 eV. Therefore, we can say that the convergence with respect to the number of $\mathbf{k}$-points is sufficiently achieved.
Although it is possible that the approximate relation that is used above does not hold for this case~\cite{ZnO_DMC}, we can expect that the convergence will be sufficient for our discussion because the finite-size error is expected to be significantly reduced in a $3\times 3\times 3$ $\mathbf{k}$-mesh from its value for a $2\times 2\times 2$ one.

\section{Imaginary part of the TC eigenvalues without the bi-orthogonal formulation}

We do not show the band structure calculated using the TC method without the bi-orthogonal extension in the main text. This is because we found that the imaginary part of the eigenvalues of the Zn-$3d$ bands obtained by solving the one-body SCF equation became as large as 0.1 eV in the TC method, which is more than 1,000 times larger than that in the BiTC method. 
Because the non-Hermiticity in the (Bi)TC formulation originates from the similarity-transformation, the presumed equivalency between the TC solution and the true many-body eigenstate is suggested to break when the imaginary parts of the eigenvalues become too large. The difference between the band structures obtained with the TC and BiTC methods was small for weakly correlated systems except LiF that is rather localized-electron system~\cite{TCMP2}. 

\section{Electron-density differences among the LDA, HF, and BiTC methods}

The differences in the electron densities calculated with the LDA, HF, and BiTC methods are shown in Fig. S1.
In Fig. S1, we can verify the tendencies explained in the main text, i.e., that the BiTC electron density tends to be localized near the atomic sites.
In addition, we can see an increase of the BiTC electron density along the Zn-O bonds near the oxygen atoms.

\begin{figure}
 \begin{center}
  \includegraphics[width=15cm]{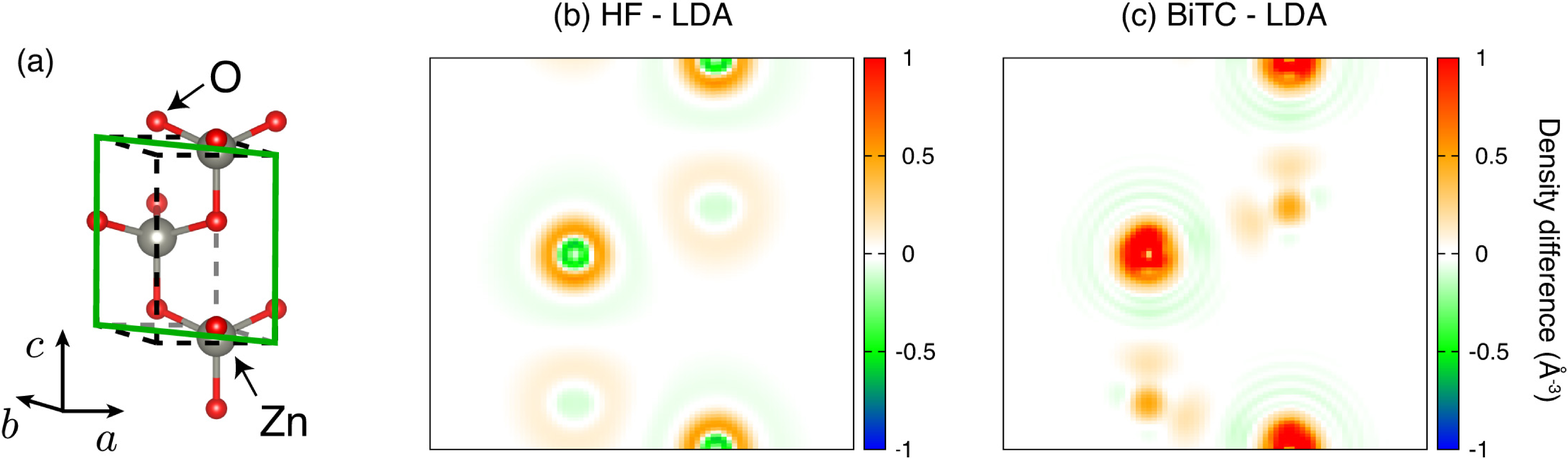}
  \caption{The differences between the HF and LDA electron densities and between the BiTC and LDA electron densities are presented in panels (b) and (c), respectively. The electron densities on the plane shown in panel (a) are depicted here.}
 \end{center}
\end{figure}


\begin{thebibliography}{999}
\bibitem{note_Vext} Note that the external potential $v_{\mathrm{ext}}$ does not include the DFT potential unlike some other post-Kohn-Sham treatments.
\bibitem{cusp_note} To fully satisfy the cusp condition for the singlet and triplet spin pairs, one should adopt the operator representation for the Jastrow function~\cite{Ten-nocusp}, which introduces the troublesome non-terminating series of the effective interaction in the TC Hamiltonian. Thus, we used this approximated cusp condition in the same manner as in many QMC studies~\cite{QMCreview,QMCreview2}.
\bibitem{note_GW} Also for weakly correlated systems, we have observed that localized states (e.g., the 3$s$ states of bulk silicon) tend to become deeper in their band structures~\cite{TCaccel,TCjfo,TCPW}.
\bibitem{note_GW2} In our observation, the similar trend is also realized for other semiconductors and insulators with $sp$ electrons, which include both the ionic and covalent crystals.
\bibitem{Tab1_note} Refs.~\cite{GWZnO} and \cite{ZnOARPES} used the lattice parameters that are a bit different from ours. However, such difference yields a difference in the band gap and the $d$-band position for less than 0.1 eV in LDA calculations. Therefore, we can conclude that the difference of the lattice parameters among several references has a minor effect on the comparison presented in Table 1.
\bibitem{note_computation} This scaling is valid only when a single Slater determinant is used (i.e. multi-determinant methodologies are not used) and the Jastrow function is represented as $u(x,x')=\sum_p w_p(x) v_{p;\sigma,\sigma'}(|\mathbf{r}-\mathbf{r}'|)w_p(x')$ (i.e. not a general two-body function).
\bibitem{note_cmptime} The ratio of computational time required for the TC and HF calculations is about 15 to 20~\cite{TCaccel}. As for the BiTC method, this ratio becomes about 20 to 40 owing to the nonequivalency of the bra and ket orbitals.
\bibitem{note_PAW} Our present implementation requires about 20,000 core hours for solving the BiTC-SCF equation, which is determined by a very large cutoff energy of plane waves. If we can make the cutoff energy an order of magnitude lower, which is a usual demand for other methods, by development of efficient core treatment, this cost will be reduced also by an order of magnitude.
\end{thebibliography}
\end{document}